\documentclass{mem}
\usepackage{graphicx} 
\usepackage{times} 
\usepackage{balance}
\usepackage{natbib} 
\usepackage{txfonts}
\usepackage{subfigure}
\usepackage[a4paper]{hyperref}
\idline{82}{3} 

\def\eg{{\it e.g.,~}}

\begin{document}

\title{Cosmic rays and diffuse non-thermal emission 
in galaxy clusters: an introduction}
\subtitle{}

\author{G. Brunetti}
\offprints{brunetti@ira.inaf.it}

\institute{INAF, Istituto di Radioastronomia, via P. Gobetti 101, 4014, 
Bologna (Italy)}

\abstract{
Recent multifrequency observations contribute to derive a 
comprehensive picture of the origin and evolution
of relativistic particles in galaxy clusters.
In this review I briefly discuss theoretical aspects and open problems of this
picture.

}

\authorrunning{G. Brunetti}
\titlerunning{Non thermal radiation in galaxy clusters}

\maketitle{}

\section{Introduction}\label{sec:intro}

Radio observations prove the existence of
non-thermal components, magnetic fields and relativistic particles, 
mixed with the hot Inter-Galactic-Medium (IGM).
The mystery of their origin rises 
new questions on the physics of the IGM.
Non-thermal components contribute to the energy of the IGM and drive physical 
processes that have the potential to
modify our present view of the IGM \cite[\eg][]{
Schekochihin2005,Subramanian2006,BL2011a}.

\noindent
Clusters host several sources of cosmic rays (CR).
CR protons are predicted to be the dominant
non-thermal particles component in the IGM (Sect. 2).
Limits on their energy content are derived from $\gamma$--rays
and radio observations (Sect. 3).

\noindent
The acceleration of relativistic electrons in the IGM
is directly probed by radio observations of diffuse (Mpc scale)
synchrotron radio emission,
{\it radio halos} and {\it relics} (e.g. Venturi, this conference).
In the last years large observational projects lead to 
significant steps in the knowledge of clusters radio properties.
The GMRT Radio Halo Survey and its combination with the 
NVSS and WENSS surveys allow a statistical exploration of
radio halos discovering a 
bimodality of the clusters's radio properties, with 
Mpc-scale radio sources found only in a fraction 
of massive clusters (Venturi, this conference). 
These surveys and their follow up in the X-rays demonstrated a tight
connection between cluster mergers and radio halos
\eg Cassano et al 2010a and ref therein).
It suggests that a fraction of the gravitational energy
that is dissipated 
during cluster mergers is channelled into the acceleration 
of relativistic particles.
Two models that connect the origin of radio halos
and relics with mergers became popular, 
turbulent and shock acceleration respectively (Sect. 4), although 
the contribution to the non-thermal emission from 
secondary particles due to proton-proton collisions in the IGM
is still debated (Sects.~3-4).

\section{CR in galaxy clusters}

\begin{figure}
\label{LrLx}
\resizebox{\hsize}{!}{\includegraphics[clip=true]{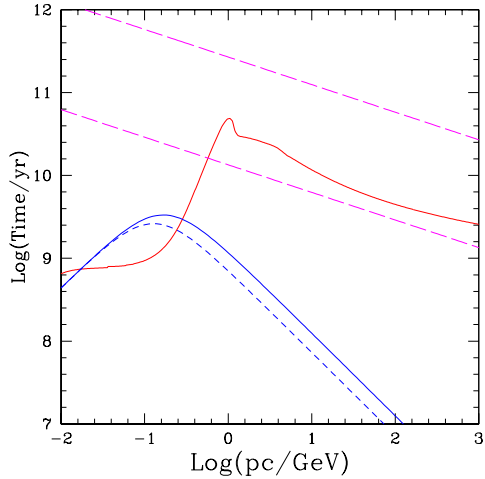}}
\caption{\footnotesize
Time scale for energy losses of protons (middle) and
electrons (bottom).
Calculations assume $z=0.1$, $n_{th}=10^{-3}$cm$^{-3}$, $B=1$ (dashed)
and $3 \mu$G (solid).
Long--dashed lines show diffusion times
assuming a Kolmogorov spectrum of magnetic
fluctuations, with $L_{30}=1$, $B=1.5 \mu$G and $\xi=$1 (upper) and
0.05 (lower).
}
\end{figure}

Particle acceleration in the IGM occurs in several places,
from ordinary galaxies
to active galaxies (AGN) \cite[\eg][]{Blasi2007}.
Nowadays, it is believed that particle acceleration at cosmological shocks
contributes the 
most to the energetics of non-thermal particles in galaxy clusters, 
with a total luminosity
in CR protons $\sim$ few $\times 10^{42} \rm erg~s^{-1}$
\cite[\eg][]{Pfrommer2006,Skillman2008,Vazza2009}.

The evolution and propagation of CR injected
in the IGM is determined by diffusion, convection and energy
losses.

\noindent
A optimistic estimate of the spatial diffusion coefficient of CR is
obtained assuming a Kolmogorov spectrum $P(k)\propto k^{-5/3}$ of
magnetic fluctuations \cite[\eg][]{Blasi2007}:

\begin{equation}
D(E) \approx 3\times 10^{29} E_{{\rm GeV}}^{1/3} \xi^{-1} B_\mu^{-1/3}
L_{30}^{2/3}~\rm cm^2 s^{-1},
\end{equation}
where $\xi$ is the fraction of the cluster magnetic field energy in 
the turbulent field and $L_{30}$ is the largest scale of magnetic 
fluctuations (in units of 30 kpc).
The corresponding time-scale to diffuse on scale $L_D=1$ Mpc,
$\tau_D \approx L_D^2 D^{-1}$, is reported in Fig.~1.
Even in the case of weakly turbulent field, $\xi = 0.05$, 
diffusion occurs in cosmological time-scales.
Faraday rotation measures (RM) suggest that a sizeable 
fraction of the magnetic field is turbulent (Murgia, this conference), 
which readily implies 
an efficient confinement of CR in galaxy clusters \citep{Voelk1996,
Berezinsky1997,Ensslin1997}.

Figure 1 also shows the time-scales for energy losses of protons
and electrons. Electrons in the IGM are short--living
particles, due to Coulomb and radiative losses.
The typical life-time is few Gyr for electrons with 
$E \approx 100$ MeV but it is much shorter, about $10^8$ yr,
for electrons with $E \approx$ few GeV, that are those 
emitting synchrotron radiation in the radio band.
Because the diffusion time on halo scales (Mpc) of 
GeV electrons is much longer than their radiative 
life-time (Fig.1), mechanisms
of in-situ (spatially distributed)
acceleration/injection of electrons are necessary to explain 
radio halos \citep{Jaffe1977}.

\noindent
Contrary to electrons, 
CR protons are long--living particles 
that can be accumulated in the cluster volume (Fig.~1). 
This leads to the conclusion 
that protons should be the dominant
non-thermal particles component in the IGM, with their
properties tracing 
the history of the complex interplay between particle acceleration
and advection processes that take place in clusters from their
formation epoch (e.g. \cite{Blasi2007} and ref. therein).

\noindent
The energy content of CR in galaxy clusters is uncertain.
Modern numerical simulations explore CR acceleration
at cosmological shocks with unprecedented details, still estimates
of $E_{CR}/E_{IGM} \approx 0.03-0.5$ unavoidably
reflect the uncertainties in the assumptions of the 
acceleration efficiency in these environments \cite[\eg][]{Ryu2003,
Pfrommer2006}.

\section{Limits on CR protons from recent observations}

\begin{figure}
\label{spettri}
\resizebox{\hsize}{!}{\includegraphics[clip=true]{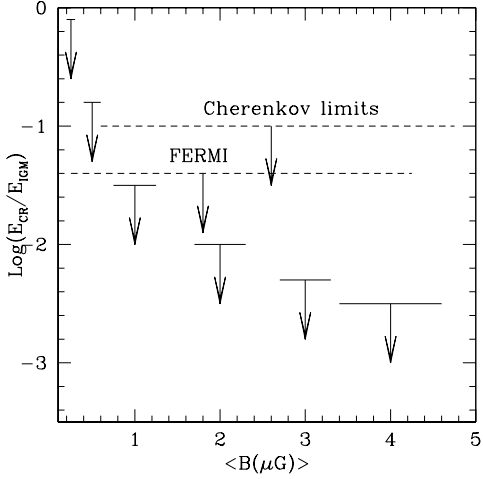}}
\caption{\footnotesize
Limits to $E_{CR}/E_{IGM}$
from Cherenkov and FERMI observations (thick--dashed
upper limit) and from radio observations (solid upper limits)
(see text).
Limits are obtained assuming
$N(p)\propto p^{-\delta}$, with $\delta = 2.2$, and that CR follow
the spatial distribution of thermal medium.
}
\end{figure}

The most direct approach to constrain the energy content of
CR protons in galaxy clusters consists in the observation of 
$\gamma$-ray emission from the decay of the neutral pions 
due to pp collisions in the IGM.

\noindent
Gamma ray upper limits from EGRET observations 
allow to put limits $E_{CR}/E_{IGM} < 0.3$ in several nearby
galaxy clusters \citep{Reimer2003}.
More stringent limits are derived from 
deep pointed observations at energies $>$100 GeV with Cherenkov telescopes.
These limits depend on 
the (unknown) spectral shape of
the proton-energy distribution. 
In the relevant case $\delta = 2$ 
($N_{CR}(p)\propto p^{-\delta}$) $E_{CR}/E_{IGM} < 0.1$ are obtained
\citep{Aharonian2009a,Aharonian2009b,Aleksic2010}, constraints
being less stringent for steeper spectra.
FERMI greatly improved the sensitivity of observations at MeV/GeV energies 
allowing a significant step.
After 18 months of observations
upper limits to the $\gamma$-ray emission of nearby clusters 
allow to derive $E_{CR}/E_{IGM} < 0.05$ \citep{Ackermann2010, Jeltema2011},
with a poor dependence on $\delta$.

Also radio observations of galaxy clusters can be used to 
put limits on $E_{CR}/E_{IGM}$ \citep{Reimer2004, Brunetti2008}.
Limits to the presence of diffuse Mpc-scale 
radio emission in clusters can be used to constrain 
secondary electrons and thus 
the energy density of the primary CR protons \citep{Brunetti2007}.
Limits from radio observations depend also on the cluster magnetic field 
strength and are complementary to those obtained from $\gamma$-rays.

Gamma and radio limits are reported in Fig.~2.
Assuming a average (Mpc-scale) magnetic field
$<B> \, \geq$ few $\mu$G, radio observations of clusters with no Mpc-scale 
radio emission provide the most stringent limits,
$E_{CR}/E_{IGM} \leq$ few$\times 10^{-3}$.
If the magnetic field is smaller, the CR energy content is
mainly constrained by FERMI.
RM provide independent constraints on the magnetic
fields of clusters suggesting
that (i) $<B> \approx 1-2 \mu$G and that (ii) clusters with and without
radio halos have similar fields (\eg Bonafede et al.~2011,
Murgia, this conference).
In this case both
radio and $\gamma$-rays imply $E_{CR}/E_{IGM} \leq$ few$\times 10^{-2}$.

\noindent
Constraints in Fig.~2 refer mainly to the innermost
($\sim$Mpc) regions of clusters where both the number density
of thermal protons (target for $\pi^o$ production) and the magnetic field
are larger. It should be mentioned that
no tight constraints are available for the clusters outskirts
where the CR contribution might be larger.

\section{A theoretical picture}

The observed connection between mergers and cluster-scale 
radio emission guides models for the origin of CR in galaxy clusters.

\noindent
Cluster mergers drive shocks in the IGM that can accelerate CR (protons
and electrons). Protons can be advected in the cluster central regions 
where they generate secondary particles trough pp collisions.
During cluster mergers large-scale turbulence is generated, 
it decays at smaller scales and can reaccelerate primary and secondary 
particles.
According to the confinement of CR, protons smoothly diffuse on Mpc
scale on clusters life-times, and guarantee a continuous source  
of secondary electrons independently of cluster dynamics.
The interplay of all these mechanisms 
implies a complex mixture of primary and secondary CR; galaxies
and AGN shall further contribute to the injection 
of primary CR in the IGM.

Radio halos and relics probe relevant aspects of this complex picture.
In the most popular (yet simplified) view 
halos are connected with cluster-turbulence (via turbulent
acceleration) and relics to cluster-shocks (via shock acceleration).
A spatial coincidence of merger shocks with edges of radio halos
(or with radio relics at the edges of halos) is now observed in
many cases \cite{Markevitch2010} suggesting that, although distinct phenomena,
relics and halos can be connected 
with the same merger.

\noindent
Hadronic models are also proposed for the origin of radio halos 
\citep[\eg][]{Blasi1999,Pfrommer2004,Keshet2010}:
extended and fairly regular radio emission from Mpc-scale
is naturally expected when the radiating electrons are 
secondaries because the parent CR protons can diffuse 
on fairly large scales.
Still several arguments may suggest that the observed giant halos 
cannot be explained by considering (``only'') hadronic collisions, e.g. :

\begin{itemize}
\item
Radio halos are very extended and most of them 
have radio brightness profiles much
broader than those predicted by hadronic models 
\citep{Dolag2000, Brunetti2004, Murgia2009,Donnert2010a,Brown2011}.
The number density of thermal
protons (targets for pp collisions) is indeed very small far  
from the cluster center implying a strong suppression of the radio brightness. 
It follows that a challenging,
very large, energy budget of CR must be postulated in this model   
to explain the extension of giant halos, at least when constraints on $B$ from
RM are taken into account (see Donnert et al 2010a,b). 
Future RM studies (e.g. eVLA) will be important.

\item 
Giant radio halos with very steep spectrum (USSRH), 
$\alpha > 1.5-1.6$ (with $F(\nu)\propto \nu^{-\alpha}$), 
exist \cite[\eg][]{Brunetti2008,
Brentjens2008,Dallacasa2009,Macario2010}, 
questioning the ``classical'' idea that 
halos have a typical spectrum $\alpha \sim 1.2$ 
(see also Venturi 2011, this conference).
Energy arguments can be used to disfavour a hadronic origin of 
USSRH \citep[\eg][]{Brunetti2004,Pfrommer2004,Brunetti2008}.
Future LOFAR observations will be {\it crucial} to test if 
these USSRH are a relevant fraction of the halos population.

\item
Gamma ray emission from galaxy clusters is unavoidably predicted by 
hadronic models \cite[\eg][]{Blasi2007}. Present FERMI upper limits 
significantly constrain the role of 
secondary electrons \citep{Ackermann2010}, challenging a hadronic
origin of several clusters-radio halos 
(\eg \cite{Jeltema2011}, see Sect.~4.3).
The improvement in sensitivity that will be achieved by FERMI in next
years will allow improving these constraints.
\end{itemize}

\subsection{Shock acceleration in galaxy clusters and
radio relics}

The morphology and polarization of radio relics suggest that their 
origin is connected with shock waves \cite[\eg][]{Ensslin1998}.
The possibility to detect relics far from cluster 
cores makes them potential probes of the process
of matter accretion in clusters.

The acceleration of CR at shocks is customarily described 
according to the diffusive shock acceleration (DSA) theory 
\cite[\eg][]{Blandford1987}
that applies when particles
can be described by a diffusion--convection equation across the shock.
For strong shocks a substantial fraction of the energy flux goes into
CR which in turn back react modifying the structure 
of shocks themselves (non linear shock acceleration theory,
\cite{Malkov1997,Blasi2002,Kang2005}).
The most relevant theoretical uncertainty
is the injection model, i.e. the probablity that
supra-thermal particles at a given velocity can leak upstream across the
subshock and get injected in the CR population.
An other major hidden ingredient is the amplification of the magnetic
field (perpendicular component) downstream
due to CR driven instabilities and adiabatic compression, as this magnetic
field self--regulates the diffusion process of supra--thermal
particles and also affects the injection process (Ryu, Kang, this 
conference).

The energetics and spectrum of the accelerated CR 
depend on the shock Mach-number. 
There is consensus on the fact that most of
the energy in clusters is dissipated at weak shocks,
with Mach numbers 2--4 (\cite{Ryu2003,Pfrommer2006, Skillman2008, Vazza2009}
see Vazza, Burns, Hoeft, this conference).
This however raises several problems :

\begin{itemize}

\item
The efficiency of CR acceleration at these weak shocks is very
low and uncertain.

\item
The acceleration of CR electrons, that is relevant for radio relics,
is still poorly understood.
\end{itemize}

\noindent
Only a handful of merger shocks have been discovered using X-ray
observations, with Mach numbers $M \approx 1.5-3$ \citep{Markevitch2001}.
Most of these shocks coincide with radio relics
or with sharp edges of radio halos indicating that 
they must have something to do with producing the observed
radio emission. Since weak shocks must be
inefficient accelerators of CR, these observations may suggest that
re-acceleration of relativistic (seeds) electrons 
is an important process in these environments (Kang, this conference).

\subsection{Turbulent acceleration \& Radio Halos}
\label{sec:dyn_cluster}

\begin{figure*}\label{LrLx}
\begin{center}
\includegraphics[width=0.99\textwidth]{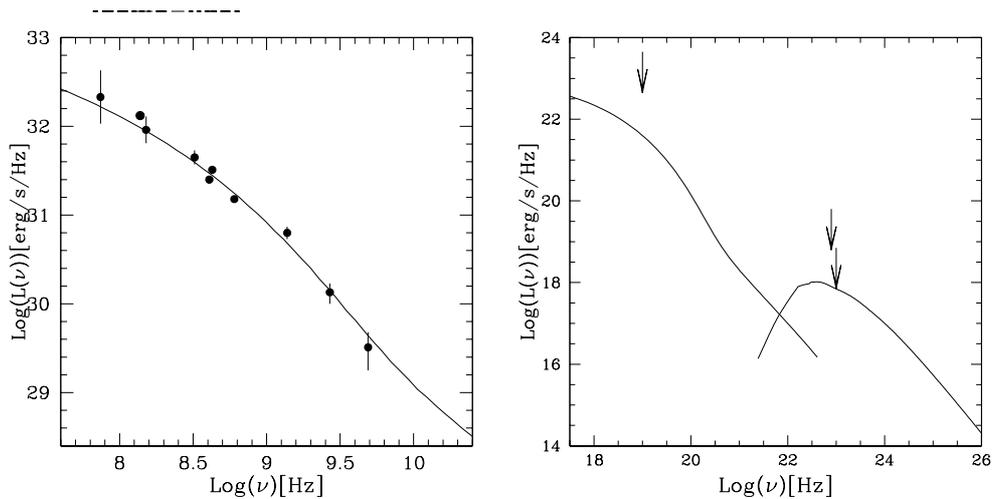}
\caption{\footnotesize
Radio and $\gamma$-ray (IC and $\pi^o$) emitted
spectrum of the Coma clusters obtained from hybrid models (readopted
from \cite{BL2011b}).
}
\end{center}
\end{figure*}

Acceleration of relativistic electrons by merger-turbulence 
is proposed for the origin of giant radio halos
\cite[\eg][]{Brunetti2001, Petrosian2001, Fujita2003}.
The ``hystorical'' motivation 
is that the particle acceleration mechanisms in radio halos are poorly
efficient as demonstrated by the steepening observed (at $\geq 1$ GHz) 
in the spectrum of the Coma halo \citep{Schlickeiser1987}.
A more recent argument in favour of turbulent acceleration
for the origin of halos comes from the discovery of
USSRH \citep[\eg][]{Brunetti2008}
that are interpreted as halos with a spectrum that steepens
at even lower radio frequencies, favouring acceleration
mechanisms poorly efficient.

Turbulence can be generated during cluster mergers 
on large scales, $L_o \sim 100-400$ kpc, with 
typical turbulent velocities
$V_o \sim 300-700$ km/s. Numerical simulations provide an unprecedented
view of this process (see Vazza, Iapichino, Jones, Paul, this conference).
Large-scale turbulence in the IGM is sub--sonic,
with $M_s = V_o/c_s \approx 0.25-0.6$, but
strongly super-Alfv\'enic (hydrodynamics), with $M_A=V_o/v_A \approx 5-10$.
At smaller scales, $l < l_A\sim L_o M^{-3}_A$, 
turbulence gets sub-Alfv\'enic (MHD) and three types of modes exist :
Alfv\'en, slow and fast modes (Lazarian, this conference).

Turbulence can accelerate particles trough resonant and non-resonant
mechanisms. 
A large part of turbulence at scales $l \geq l_A$
could be in the form of compressible motions and 
Transit-Time-Damping is the most relevant mechanism \citep[\eg][]{BL2011b}.

At smaller scales, $l \leq l_A$, most of turbulence is probably in the 
form of Alfv\'en and slow modes whose contribution to particle
acceleration via resonant (gyro-resonant) interaction is potentially
strong but more uncertain due to the anisotropy that develops when
these modes cascade from larger to smaller scales \citep[\eg][]{Yan2004}.
Consequently Alfv\'enic models must postulate 
the injection of Alfv\'en modes ``directly'' at small scales.
This may occur trough several processes in the IGM (Lazarian, this
conference) although self-consistent calculations are not yet available.

Several open questions exist in turbulent acceleration models, e.g. :

\begin{itemize}

\item
The efficiency of turbulent acceleration in the IGM is still
poorly constrained, due to the fact that a self-consistent model of the
multi-scale turbulence in galaxy clusters
is still missing (\cite{BL2011a} for recent attempts).

\item
Theoretically this model provides a ``unique'' avenue to explain
the {\it large spatial extent of giant radio halos} and the {\it variety of
the observed spectral properties of halos}.
However this has not be proved yet 
by detailed numerical simulations (see however
first efforts in this direction, Donnert, ZuHone, 
this conference).

\end{itemize}

Although the uncertainties in the details of this complex model, 
future observations will allow clear tests.
The model predicts the existence of USSRH \citep[\eg][]{Cassano2006}. 
This is the consequence of the fact that spectral steepening occurs at
lower frequencies in those halos that are generated in less energetic
merger events (see Cassano, this conference).
Radio halos with very--steep spectrum must
be common in this scenario and LOFAR surveys can test 
this \citep{Cassano2010b}.

\subsection{``Hybrid'' models and constraints from $\gamma$-ray limits}

Acceleration of electrons from the thermal pool to relativistic energies 
by MHD turbulence in the IGM faces serious problems due to energy arguments 
\citep[\eg][]{Petrosian2008}.
Consequently, turbulent acceleration models assume 
a pre-existing population of relativistic particles that
provides the seeds to ``reaccelerate'' during mergers.

\noindent
In ``hybrid'' models {\it seeds are secondary electrons} \citep{BB2005}. 
Recent calculations have demonstrated that reacceleration of CR and
their secondaries
may be sufficient to generate the observed radio halos in merging clusters, 
including their {\it observed brightness profiles and spectra}, provided that 
the spatial distribution of the primary CR is relatively flat \citep{BL2011b}. 
Expectations are 
consistent with present upper limits (FERMI) provided that the
magnetic field in clusters is in agreement with 
that derived from RM (Fig.~4).

\noindent
Having in hands a complete treatment of primary and 
secondary particles in the IGM, ``hybrid'' calculations emulate 
hadronic models 
in the case we assume ``no'' (or negligible) turbulence.
In this case the energy of CR that is necessary to reproduce the
radio luminosity 
and the observed brightness profile of radio halos is much larger 
than that in the case of turbulent reacceleration; e.g. for 
Coma it is about 10 times larger than that in Fig.~4 
($E_{CR}/E_{IGM} \approx 0.3$, for $\delta =2.6$, considering the 
magnetic field from \cite{Bonafede2010}).
The consequence, {\it 
in addition to the well known difficulty to reproduce the 
radio spectrum}, is a $\gamma$-ray flux 
that becomes inconsistent with the present FERMI limit (Fig.~5), unless 
$B$ is 2 times larger than that from RM.

\begin{figure*}\label{LrLx}
\begin{center}
\includegraphics[width=0.99\textwidth]{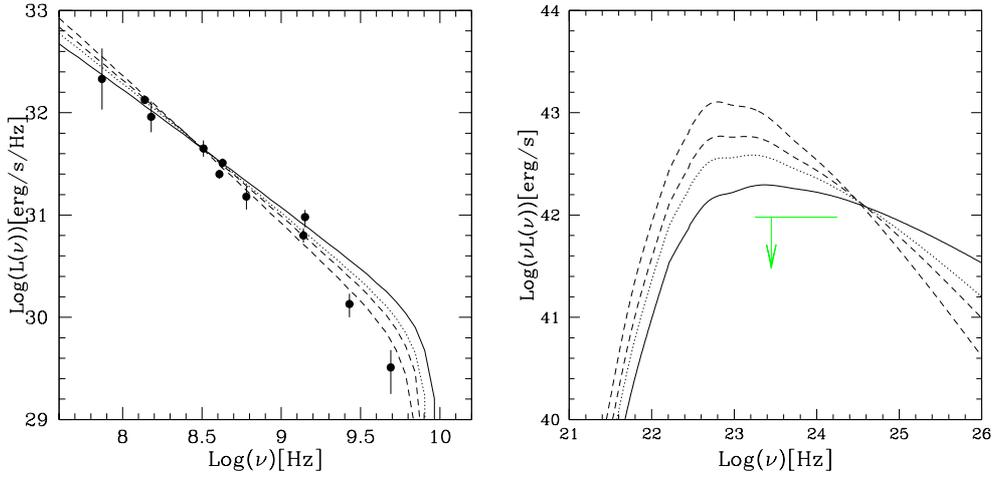}
\caption{\footnotesize
Radio (including SZ decrement) and $\gamma$-ray ($\pi^o$) 
spectrum of the Coma clusters from secondary models assuming a magnetic
field from \cite{Bonafede2010} and the observed brightness profile
from \cite{Govoni2001}. Different models assume $\delta=2.6\, .. 3.2$.}
\end{center}
\end{figure*}

\section{Evolution of radio halos}

It was quickly realised that radio halos
are not common \citep{Giovannini1999} and that this is 
a challenge for a hadronic models which predict long-living halos in all
clusters \citep{Hwang2004}.
Recent surveys, such as the GMRT Radio Halo 
Survey (RHS, \cite{Venturi2008}), start the study 
of the occurrence of
halos in clusters and their evolution with cosmic time. 
The RHS allows the discovery of a bimodal behaviour
of galaxy clusters : in a $P_{1.4}$--$L_X$ diagram ``radio quiet''
clusters are well separated from giant radio halos (Fig.~5).
When combined with the radio halo -- merger connection, 
the bimodality suggests that (i) relativistic particles are
accelerated in mergers and cool when systems becomes more 
relaxed \citep{Brunetti2007,Brunetti2009},
or (ii) that the magnetic field is amplified during mergers
and dissipated in relaxed clusters \citep{Kushnir2009,Keshet2010}.
Present analysis of RM in clusters' radio sources 
do not find a difference between the 
magnetic properties of clusters with and without
halos, supporting the scenario where the radio bimodality
is due to merger-induced acceleration mechanisms
(\cite{Bonafede2011} and ref. therein).

\noindent
It has been claimed
that if CR propagate at super-Alfv\'enic speeds in (less
turbulent) clusters, propagation of CR can also induce
radio bimodality \citep{Ensslin2011}.
Super-Alfv\'enic propagation of CR however faces difficulties   
as it would get quenched by the scattering by the MHD waves that are 
generated by CR streaming 
itself (\cite{Schlickeiser2002} for a modern review including 
high beta plasmas, such as the IGM).

\begin{figure}
\label{spettri}
\resizebox{\hsize}{!}{\includegraphics[clip=true]{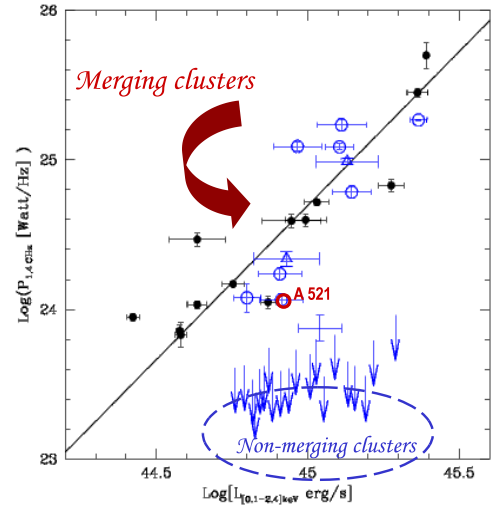}}
\caption{Distribution of galaxy clusters in the $P_{1.4}-L_X$ diagram
(readapted from \cite{Brunetti2009}).}
\end{figure}

\section{Conclusions}

Recent observations lead to a substantial progress in constraining
the energy content of CR in clusters and the properties of Mpc-scale
radio sources associated with a fraction of merging clusters.
Present view of the non-thermal particle
components in the IGM is very complex with several main players, 
shocks, turbulence and hadronic collisions.
It is believed that radio halos and relics probe CR in turbulent and 
shocked regions, although
the details of the mechanisms responsible for their origin 
are still unclear and observations with future radiotelescopes, such as 
LOFAR, will be crucial to test present models.

\noindent
Although radio observations, and their follow up in the X-rays, 
contribute the most to our understanding of
non-thermal phenomena in galaxy clusters,
an important contribution is now coming from $\gamma$-ray
observations. They put stringent limits on the energy content
of CR in clusters and also provide new
challenges for a ``pure'' hadronic origin of radio halos when combined
with independent information on clusters magnetic fields.

\begin{acknowledgements}
This work is partially supported by INAF under grants PRIN-INAF08 and 09.
\end{acknowledgements}

\bibliographystyle{aa}

\begin{thebibliography}{}

\bibitem[Aleksic et al. (2010)]{Aleksic2010} 
Aleksic J., et al. 2010, \apj, 710, 634

\bibitem[Ackermann et al.(2010)]{Ackermann2010} 
Ackermann M., et al. 2010, \apj, 717, L71

\bibitem[Aharonian et al.(2009a)]{Aharonian2009a} 
Aharonian F.A., et al., 2009a, A\&A, 495, 27

\bibitem[Aharonian et al.(2009b)]{Aharonian2009b}
Aharonian F.A., et al., 2009b, A\&A, 502, 437

\bibitem[Bonafede et al. (2010)]{Bonafede2010} 
Bonafede A., et al., 2010, A\&A, 513, 30

\bibitem[Bonafede et al. (2011)]{Bonafede2011} 
Bonafede A., et al., 2011, arXiv:1103.0277

\bibitem[Berezinsky et al. (1997)]{Berezinsky1997}
Berezinsky V.S., et al., 1997, \apj, 487, 529

\bibitem[Blandford \& Eichler (1987)]{Blandford1987}
Blandford R., Eichler D., 1987, PhR, 154, 1

\bibitem[Blasi (2002)]{Blasi2002}
Blasi P., 2002, APh, 16, 429 

\bibitem[Blasi\& Colafrancesco (1999)]{Blasi1999}
Blasi P., Colafrancesco S., 1999, APh, 12, 169

\bibitem[Blasi et al. (2007)]{Blasi2007}
Blasi P., et al., 2007, IJMPA 22, 681

\bibitem[Brentjens (2008)]{Brentjens2008}
Brentjens M.A., 2008, A\&A, 489, 69

\bibitem[Brown \& Rudnick (2011)]{Brown2011}
Brown S., Rudnick L., 2011, \mnras, 412, 2

\bibitem[Brunetti (2004)]{Brunetti2004}
Brunetti G., 2004, JKAS, 37, 493

\bibitem[{Brunetti et al. (2001)}]{Brunetti2001}
Brunetti, G., et al., 2001, MNRAS, 320, 365

\bibitem[{Brunetti \& Blasi (2005)}]{BB2005}
Brunetti, G., Blasi P., 2005, \mnras, 363, 1173 

\bibitem[{Brunetti et al. (2007)}]{Brunetti2007}
Brunetti, G., et al.,
2007, ApJ, 670, L5

\bibitem[{Brunetti et al. (2008)}]{Brunetti2008}
Brunetti, G., et al.,
2008, Nat, 455, 944

\bibitem[{Brunetti et al. (2009)}]{Brunetti2009}
Brunetti, G., et al.,
2009, A\&A, 507, 661

\bibitem[{Brunetti \& Lazarian (2011a)}]{BL2011a}
Brunetti G., Lazarian A., 2011a, \mnras, 412, 817

\bibitem[{Brunetti \& Lazarian (2011b)}]{BL2011b}
Brunetti G., Lazarian A., 2011b, \mnras, 410, 127

\bibitem[Cassano et al. (2006)]{Cassano2006} 
Cassano, R., et al.
2006, \mnras, 369, 1577

\bibitem[{Cassano et al. (2010a)}]{Cassano2010a}
Cassano, R., et al.,
2010a, ApJL, 721, L82

\bibitem[{Cassano et al. (2010b)}]{Cassano2010b}
Cassano, R., et al.,
2010b, A\&A, 509, 68

\bibitem[Dallacasa et al. (2009)]{Dallacasa2009}
Dallacasa D., et al., 2009, \apj, 699, 1288

\bibitem[Dolag \& Ensslin (2000)]{Dolag2000} 
Dolag K., En\ss lin T.A., 2000, A\&A, 362, 151

\bibitem[Donnert et al. (2010a)]{Donnert2010a} 
Donnert J., et al., 2010a, \mnras, 407, 1565

\bibitem[Donnert et al. (2010b)]{Donnert2010b} 
Donnert J., et al., 2010b, \mnras, 401, 47

\bibitem[En\ss lin et al. (1997)]{Ensslin1997} 
En\ss lin T.A., et al., 1997, \apj, 477, 560

\bibitem[En\ss lin et al. (1998)]{Ensslin1998}
En\ss lin T.A., et al., 1998, A\&A, 333, 47

\bibitem[En\ss lin et al. (2011)]{Ensslin2011}
En\ss lin T.A., et al., 2011, A\&A, 527, 99

\bibitem[Fujita et al. (2003)]{Fujita2003} 
Fujita Y., et al., 2003, \apj, 
584, 190

\bibitem[Giovannini et al. (1999)]{Giovannini1999} 
Giovannini, G., et al., 1999, NewA, 4, 141

\bibitem[Govoni et al. (2001)]{Govoni2001} 
Govoni, F., et al., 2001, A\&A, 369, 441

\bibitem[Hwang (2004)]{Hwang2004} 
Hwang C.-Y., 2004, JKAS, 37, 461

\bibitem[Jaffe (1977)]{Jaffe1977}
Jaffe W.J., 1977, \apj, 212, 1

\bibitem[Jeltema \& Profumo (2011)]{Jeltema2011}
Jeltema T.E., Profumo S., 2011, \apj, 728, 53

\bibitem[Kang \& Jones (2005)]{Kang2005}
Kang H., Jones T.W., 2005, \apj, 620, 44

\bibitem[Keshet \& Loeb (2010)]{Keshet2010}
Keshet R., Loeb A., 2010, \apj, 722, 737

\bibitem[Kushnir et al. (2009)]{Kushnir2009}
Kushnir, D., et al., 2009, JCAP, 9, 24

\bibitem[{Macario et al. (2010)}]{Macario2010}
Macario, G., et al.,
2010, A\&A, 517, A43

\bibitem[Malkov (1997)]{Malkov1997}
Malkov M.A., \apj, 485, 638

\bibitem[{Markevitch (2010)}]{Markevitch2010}
Markevitch, M., 2010, arXiv:1010.3660

\bibitem[{Markevitch \& Vikhlinin (2001)}]{Markevitch2001}
Markevitch, M., \& Vikhlinin, A.\ 2001, \apj, 563, 95

\bibitem[{Murgia et al. (2009)}]{Murgia2009}
Murgia, M., et al., 2009, A\&A, 499, 679 

\bibitem[Petrosian (2001)]{Petrosian2001} 
Petrosian V., 2001, \apj, 557, 560 

\bibitem[Petrosian \& East (2008)]{Petrosian2008} 
Petrosian V., East W.E., 2008, \apj, 682, 175

\bibitem[Pfrommer \& Ensslin (2004)]{Pfrommer2004} 
Pfrommer C., Ensslin T.A., 2004, \mnras, 352, 76

\bibitem[Pfrommer et al. (2006)]{Pfrommer2006} 
Pfrommer C., et al., 2006, \mnras,
367, 113

\bibitem[Reimer et al. (2004)]{Reimer2004} 
Reimer A., et al., 2004
A\&A 424, 773

\bibitem[Reimer et al. (2003)]{Reimer2003} 
Reimer O., et al., 2003, \apj, 588, 155

\bibitem[Ryu et al. (2003)]{Ryu2003} 
Ryu, D., et al., 2003, \apj, 593, 599

\bibitem[Schekochihin et al. (2005)]{Schekochihin2005} 
Schekochihin A.A., et al., 2005, \apj, 629, 139

\bibitem[Schlickeiser et al. (1987)]{Schlickeiser1987} 
Schlickeiser R., et al., 1987, A\&A, 182, 21

\bibitem[Schlickeiser (2002)]{Schlickeiser2002} 
Schlickeiser R., 2002, {\it Cosmic Ray Astrophysics}, Springel

\bibitem[Skillman et al. (2008)]{Skillman2008} 
Skillman S.W., et al., 2008, \apj, 689, 1063

\bibitem[Subramanian et al. (2006)]{Subramanian2006} 
Subramanian K., et al., 2006, \mnras, 366, 1437

\bibitem[Vazza et al. (2009)]{Vazza2009} 
Vazza F., et al., 2009, \mnras, 395, 1333

\bibitem[{Venturi et al. (2008)}]{Venturi2008}
Venturi, T., et al.,
2008, A\&A, 484, 327

\bibitem[Voelk et al. (1996)]{Voelk1996} 
V\"{o}lk H.J., et al., 1996, SSRv, 75, 279

\bibitem[Yan \& Lazarian (2004)]{Yan2004}
Yan, H., Lazarian, A., 2004, \apj, 614, 757

\end{thebibliography}

\end{document}